\documentclass[aps,floats,floatfix,superscriptaddress]{revtex4-2}

\newcommand{\beq}{\begin{equation}}
\newcommand{\eeq}{\end{equation}}

\newcommand{\br}{{\bf r}}

\usepackage[english]{babel}
\usepackage {indentfirst}
\usepackage{comment} 
\usepackage[letterpaper,top=2cm,bottom=2cm,left=2.25cm,right=2.25cm,marginparwidth=1.75cm]{geometry}
\usepackage{amsmath,amssymb,amsthm}
\usepackage{graphicx}
\usepackage{float}
\usepackage{bm}  
\usepackage{hyperref}
\hypersetup{
    colorlinks=true,
    linkcolor=blue,
    filecolor=blue,
    urlcolor=blue,
    citecolor=blue,
}
\usepackage{nicefrac}

\newcommand{\diag}{\mathrm{diag}}

\usepackage[normalem]{ulem}

\usepackage[mathlines]{lineno}  
\begin{document}
\title{
Rapid general Electromagnetic Analysis with computational conformal geometry via Conformal Energy Minimization }
\author{Pengcheng Wan}
\affiliation{Shanghai Institute for Mathematics and Interdisciplinary Sciences, Shanghai 200433, China}
\affiliation{ Research Institute of Intelligent Complex Systems, Fudan University, 200433 Shanghai, China }

\author{Zhong-Heng Tan}
\affiliation{School of Mathematics and Shing-Tung Yau Center, Southeast University, Nanjing 210096, China}

 \author{S. T. Chui}
\altaffiliation{Corresponding authors. \\ 
Email: \href{mailto:[your.email@example.com]}{chui@udel.edu}; \href{mailto:[your.email@example.com]}{txli@seu.edu.cn}; \href{mailto:[your.email@example.com]}{styau@tsinghua.edu.cn}}
 \affiliation{Bartol Research Institute and Department of Physics and Astronomy, University of Delaware, Newark, DE 19716, USA}
 \affiliation{Shanghai Institute for Mathematics and Interdisciplinary Sciences, Shanghai 200433, China}

\author{Tiexiang Li}
\altaffiliation{Corresponding authors. \\ 
Email: \href{mailto:[your.email@example.com]}{chui@udel.edu}; \href{mailto:[your.email@example.com]}{txli@seu.edu.cn}; \href{mailto:[your.email@example.com]}{styau@tsinghua.edu.cn}}
\affiliation{School of Mathematics and Shing-Tung Yau Center, Southeast University, Nanjing 210096, China}
\affiliation{Shanghai Institute for Mathematics and Interdisciplinary Sciences, Shanghai 200433, China}

\author{ S. T. Yau}
\altaffiliation{Corresponding authors. \\ 
Email: \href{mailto:[your.email@example.com]}{chui@udel.edu}; \href{mailto:[your.email@example.com]}{txli@seu.edu.cn}; \href{mailto:[your.email@example.com]}{styau@tsinghua.edu.cn}}
 \affiliation{\mbox{Yau Mathematical Sciences Center, Jingzhai, Tsinghua University, Haidian District, Beijing 100084, China}} 


\begin{abstract}
We recently found that the electromagnetic scattering problem can be very fast in an approach expressing the fields in terms of orthonormal basis functions. In this paper we apply computational conformal geometry with the conformal energy minimization (CEM) algorithm to make possible fast solution of finite-frequency electromagnetic problems involving {\bf arbitrarily shaped}, simply-connected metallic surfaces. The CEM algorithm computes conformal maps with minimal angular distortion, enabling the transformation of arbitrary simply-connected surfaces into a disk, where orthogonal basis functions can be defined and electromagnetic analysis can be significantly simplified. We demonstrate the effectiveness and efficiency of our method by investigating the resonance characteristics of two metallic surfaces: a square plate and a four-petal plate. Compared to traditional finite element methods (e.g., COMSOL), our approach achieves a three-order-of-magnitude improvement in computational efficiency, requiring only seconds to extract resonant frequencies and fields. Moreover, it reveals low-energy, doubly degenerate resonance modes that are elusive to conventional methods. These findings not only provide a powerful tool for analyzing electromagnetic fields on complex geometries but also pave the way for the design of high-performance electromagnetic devices.
\end{abstract}
\maketitle

\vspace{-2.0em} 
\section{Introduction}

The electromagnetic scattering from arbitrarily shaped metallic surfaces constitutes an almost ubiquitous and critical problem, spanning applications from radar cross-section analysis\cite{knott2004radar} to metasurface\cite{yu2014flat,brongersma2025second} and sensor\cite{chui2017electromagnetic} design. Its core lies in the rapid decoding of fundamental wave-matter interactions, particularly in probing the complex dynamic mechanisms underlying electromagnetic wave manipulation and the resonant behavior of artificial microstructures\cite{pendry2000negative,smith2004metamaterials,chen2006active,yu2011light}. This problem involves both theoretical modeling and the design of efficient algorithms. However, 
finite element methods are too slow for real-time industrial applications and exhibit convergence instability and exponentially escalating computational costs at geometric boundaries like sharp edges and corners.

Recently we have  developed an efficient and rigorous approach by representing the fields and the Green's function, which corresponds to the capacitance and the inductance, in terms of complete orthornormal basis functions\cite{chui2012electromagnetic,zhou2006eigenmodes,zhan2014t,zhan2015multiple}.  A simple and effective method for handling boundary conditions is obtained by introducing electric fields at the boundary, the value of which can be self-consistently determined\cite{chui2014resonances,chui2017electromagnetic,cone1}.  
For the examples we have studied, our approach is about a thousand times faster than finite element methods with comparable accuracy. For the geometry of a cone with a sharp tip that corresponds to the atomic force microscope tip\cite{cone1},
our approach agrees well with experimental results for near field scattering microscopy whereas results with finite elements methods are off by 20 per cent.

An obstacle to applying our approach for arbitrary surfaces is that the orthonormal basis functions are generally not known. This problem can be solved with computational conformal geometry\cite{gu2008computational}, with numerical algorithms for achieving conformal harmonic mappings between simple and complex surfaces.  Orthonormal basis functions for complex surfaces can be obtained by this map from the known basis functions of simple surfaces. In this paper we demonstrated this approach. With the conformal energy minimization (CEM) \cite{yueh2017efficient, YCWW21} algorithm we generate basis functions for two examples of different geometries, a square plate and a four-petal plate. The four-petal shape is formed by attaching a semicircle with a diameter equal to the edge length to each side of a square.
We then compute the Green’s function representation using these orthonormal basis functions and solve finite-frequency electromagnetic problems. The results demonstrate excellent agreement with those obtained from traditional finite element methods.
The extraction of these physical quantities can be completed within seconds, outperforming conventional finite element methods (e.g., COMSOL) by three orders of magnitude in computational efficiency for resonant frequency determination.

The CEM algorithm compute conformal maps with minimal angular distortions by combining rigorous energy minimization with practical computational optimizations. Moreover, it handles surfaces with large numbers of mesh points in seconds, enabling real-time applications in our approach.  The calculated results for the square plate with the conformal mapping obtained by the CEM agree with those using the analytic conformal mapping via the Schwarz-Christoffel (SC) transformation, thereby validating the CEM algorithm. Our method also revealed low-energy doubly degenerate resonance modes that are elusive to traditional methods.
These findings provide an efficient toolbox for addressing electromagnetic field problems involving complex geometric metallic surfaces, and hold significant implications for the design of high-performance electromagnetic devices. We now describe our results in detail.

 \vspace{-0.5em}
\section{Solution of Maxwell's equation}
\begin{figure}[tbph]%
\vspace*{0pt} \centerline{\includegraphics[angle=0,width=8cm]{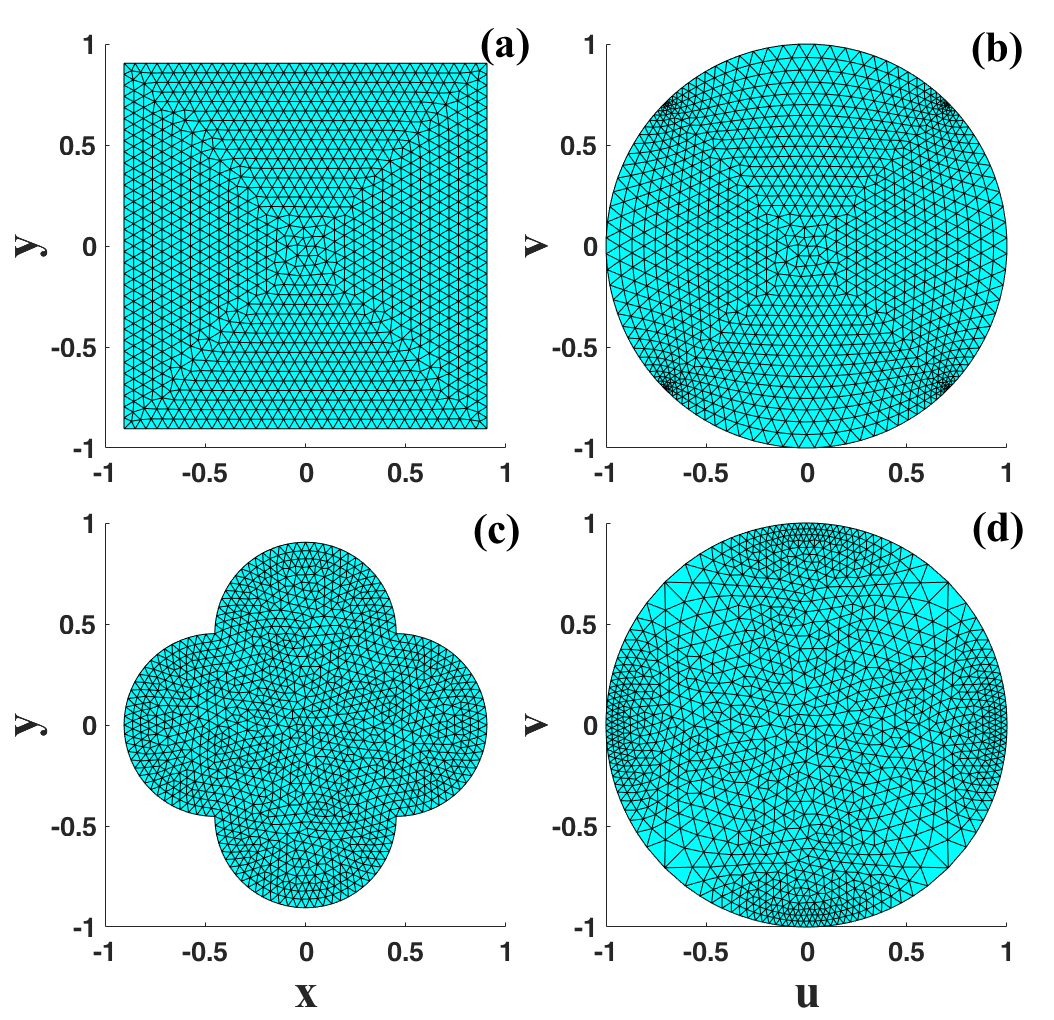}}
\vspace*{0pt}
\caption{{\bf Discrete conformal maps computed using the CEM algorithm.} Discrete conformal maps between figures on the left and those on the right:
(a) the square plate to (b) the unit disk (N = 2,490 triangular elements);
(c) the four-petal plate to (d) the unit disk (N = 2,488 triangular elements). Both plates have an edge length of 1.81.}
\label{fig:diskmaps}
\end{figure}

In this section, we describe our approach to solving the electromagnetic problem in terms of an orthonormal basis set.
To investigate the response of a metallic surface to an external electromagnetic(EM) field, we assume that the physical quantities are expressed in terms of their Fourier components in time dependence $e^{i\omega t}$ with angular frequency $\omega$. Unless explicitly stated, we shall solve the equations for each Fourier component. The current density $\mathbf{j}$ under the influence of the total external EM field $\mathbf{E}^t_{\text {ext }}$ 
is governed by the equation
\beq
\rho_0 \mathbf{j}= \mathbf{E}_{\mathrm{em}}+\mathbf{E}^t_{\mathrm{ext}},
\label{j0}
\eeq
where $\rho_0$ is the resistivity, $\mathbf{E}_{\text {em }}$ is the electromagnetic field generated by the current. 
The total external  field
\beq
\mathbf{E}^t_{\mathrm{ext}}=\mathbf{E}_{\mathrm{ext}}+\mathbf{E}_{s}
\label{be}
\eeq
 is a sum of the externally applied electric field $\mathbf{E}_{\mathrm{ext}}$ and a boundary field $\mathbf{E}_s$ from the boundary condition and which is self-consistently determined.
$\mathbf{E}_{\text {em }}$ of a metallic surface can be represented as $\mathbf{E}_{\text {em }}=-\mathbf{Z}^{\mathbf{0}} \mathbf{j}$. Eq. (\ref{j0}) can be written as
 \begin{equation}
\mathbf{j}=\mathbf{Z}^{-1}\mathbf{E}^t_{\mathrm{ext}}, 
 \label{eq:Cir}
\end{equation}
where $\mathbf{Z}=\mathbf{Z}^{0}+\mathbf{1} \rho_{0}$. 
The ``impedance'' matrix $\mathbf{Z}^{0}$ is just the representation of the Green's function in an orthonormal set of basis functions. 
The basis functions can be obtained with conformal mapping.  We illustrate the conformal mapping for two examples shown in Figure \ref{fig:diskmaps}.  

Each surface we considered is represented as a triangulation mesh with N elements in a certain order. Based on the CEM algorithm, we obtain a discrete conformal map that maps a point $z=x+i y$ in the surface  of the left to a point  $w=u+i v$ in the corresponding circle on the right, as shown in Figure \ref{fig:diskmaps} (Between \ref{fig:diskmaps}(a) and \ref{fig:diskmaps}(b); \ref{fig:diskmaps}(c) and \ref{fig:diskmaps}(d)). With this kind of mapping, we construct the orthonormal vector basis functions $\mathbf{c M}[w(z)]$ and $\mathbf{c N}[w(z)]$ for the surface, respectively
\begin{equation}
\begin{aligned}
& \mathbf{c M}_{m}[kw(z)]=\left(p_{+} f_{m+1} \mathbf{e}_{-}+p_{-} f_{m-1} \mathbf{e}_{+}\right) / 2,  \\
& \mathbf{c N}_{m}[kw(z)]=\left(-p_{+} f_{m+1} \mathbf{e}_{-}+p_{-} f_{m-1} \mathbf{e}_{+}\right) / 2,
\end{aligned}
\label{eq:cMcN}
\end{equation}
where $p_{+}=d w^{*} / d z^{*}, p_{-}=d w / d z$ are the derivatives of the conformal map, $\mathbf{e}_{\pm}=\mathbf{e}_{x} \pm i \mathbf{e}_{y}$ are circularly polarized unit vectors, $f_{m \pm 1}(kw(z))=\exp [i(m \pm 1) \phi_{w}(z)] J_{m \pm 1}(k|w(z)|) / C_{m}$; $J_{m}$ is the Bessel function with the angular momentum index $m$ , $k$ is the wave number. The normalization constant is given by $C_{m}^2=\int J^{2} r d r=$ $J^{2}\left(R^{2}-m^{2} / k^{2}\right) / 2+R^{2} J^{\prime 2} / 2$, where $R$ is the radius of the circular disk. The discrete set of wave numbers $\{k\}$ are determined by either the ``V'' (for ``value'') basis, $J_{m}(k R)=0$, or the ``D'' (for ``derivative'') basis, $J_{m}^{\prime}(k R)=0$. Thus we can expand the current on the surface in terms of our basis
\begin{equation}
\mathbf{j}=\sum_{m,\kappa}\{j_{M, m, \kappa} \mathbf{cM}_{m}\left[k_{\kappa}w(z)\right]+j_{N, m, \kappa} \mathbf{cN}_{m}\left[k_{\kappa}w(z)\right]\},
\label{eq:jcMN}
\end{equation}
where the subscripts $M, N$ represent the components of the basis functions $\mathbf{cM}, \mathbf{cN}$ respectively, $\kappa$ denotes the index for the discrete set of wave numbers $\{k\}$.

The boundary field from Eq.\eqref{be} can be expressed in terms of the variables $w$ in the circle(with radius $R$) as $E_{S}\left(\phi_{w}\right)=\sum_{l} E_{S}(l) \exp \left(i l \phi_{w}\right) /(2 \pi)$. They are determined from the boundary condition that the normal boundary currents become zero. In terms of our basis functions $\mathbf{I}=\mathbf{cM}, \mathbf{cN}$, Eq. \eqref{eq:Cir} becomes
\begin{equation}
j_{I, m, \kappa}=\sum_{m^{\prime}, \kappa^{\prime}, I^{\prime}=M,N} Z_{I, m, \kappa ; I^{\prime}, m^{\prime},\kappa^{\prime}}^{-1} \cdot\left[E_{\text {ext },  I^{\prime}, m^{\prime}, \kappa^{\prime}}+\sum_{l} B_{I^{\prime}}\left(m^{\prime}, l, \kappa^{\prime}\right) E_{S}(l)\right],
\end{equation}
here
\begin{equation}
B_{I}(m,n,\kappa)=\int d z d z^{*} \exp \left(i n \phi_{w}\right) \delta(|w(z)|-R) \mathbf{e}_{R}^{z} \cdot \mathbf{I}_{m}^{*}\left[k_{\kappa}w(z)\right]
\label{eq:Bint}
\end{equation}
transforms the radial component of basis function at the edge for the finite surface to that of the circle, where $\mathbf{e}_{R}^{z}$ is perpendicular to the perimeter of the  finite surface. The integral is taken over the variables on the boundary.

The boundary condition is that the radial component of the current
 $j_{r}\left(|w(z)|=R\right)=0$, for all values of $\phi$.

From this, we finally get
\begin{equation}
\mathbf{E}_{s}=-\mathbf{B} \mathbf{K}^{-1} \mathbf{B Z}^{-1} \mathbf{E}_{\mathrm{ext}},  \ \mathbf{K}=\mathbf{B} \mathbf{Z}^{-1} \mathbf{B},
\label{eq:Es&K}
\end{equation}
where $\mathbf{K}$ is the inverse impedance $\mathbf{Z}^{-1}$ projected onto the angular momentum basis of the finite boundary:
\begin{equation}
K_{n, l}=\sum_{m, m^{\prime}, \kappa, \kappa^{\prime}, l, I^{\prime}} B_{I}(n, m, \kappa) Z_{I, m, \kappa ; I^{\prime}, m^{\prime}, \kappa^{\prime}}^{-1} B_{I^{\prime}}\left(m^{\prime}, l, \kappa^{\prime}\right).
\label{eq:Knl}
\end{equation}

The resonance condition can come when the boundary surface field $\mathbf{E}_{s}$ or the response to the ``raw'' external field  is divergent:
\begin{equation}
\operatorname{det}(\mathbf{K})=0 \text {, or } \operatorname{det}(\mathbf{Z})=0 \text {. }
\label{eq:detK}
\end{equation}

Combining the above Eqs \eqref{eq:Cir} and \eqref{eq:Es&K}, we get
\begin{equation}
\mathbf{j}=\left[1-\mathbf{Z}^{-1} \mathbf{B} \mathbf{K}^{-1} \mathbf{B}\right] \mathbf{Z}^{-1} \mathbf{E}_{\mathrm{ext}} .
\end{equation}
Eqs. \eqref{eq:Cir}, \eqref{eq:Es&K} and \eqref{eq:detK} constitute the core results of our approach. The remaining task is the numerical calculation of the Green's/Impedance function matrix  $\mathbf{Z}$ and its inversion. As is previously mentioned, this matrix is nearly diagonal with off-diagonal elements rapidly decaying. The usual interest lies in the low-lying modes, thus only a few terms are required for an accurate result.
\section{Results and Discussions}
\subsubsection{Validation of our numerical conformal mapping}
\begin{figure}[tbph]
\vspace*{0pt} \centerline{\includegraphics[angle=0,width=8cm]{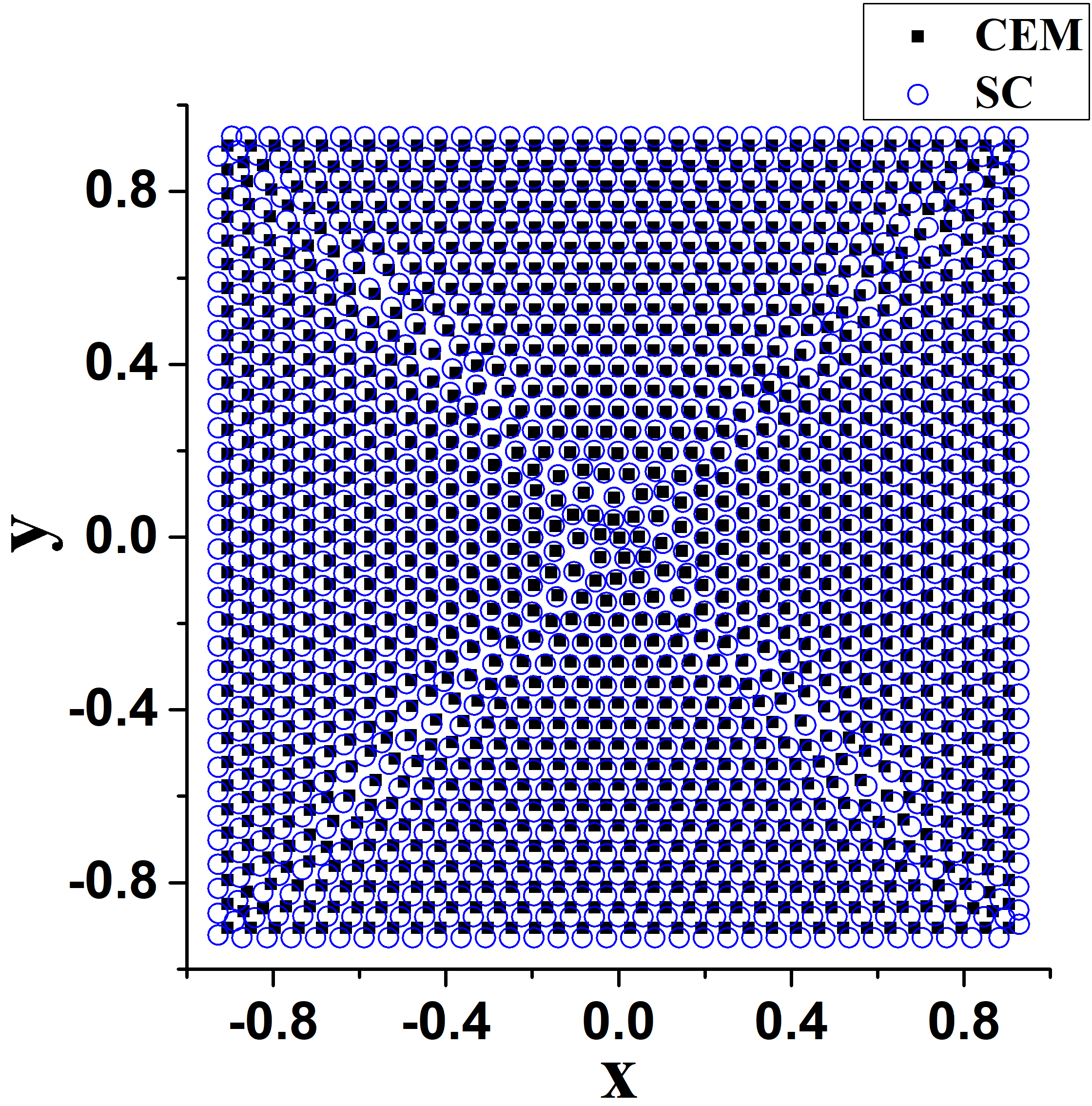}}
\vspace*{0pt}
\caption{{\bf Comparison of numerical and analytical conformal mappings.} Triangular mesh vertices obtained by numerical (CEM) and analytical (SC) methods for mapping the unit disk to the square.}
\label{val}
\end{figure}

For the regular polygon with K sides, an analytic form of the conformal mapping is given by the Schwarz-Christoffel transformation
\beq
z = S(w) = A\int_0^w dw'G, 
\label{scz}
\eeq
where $G=\Pi_{k=1}^K (1-w'\exp(-i 2\pi k/K)^{-2/K},$ $A$ is an arbitrary constant that determines the scale of the transformation.
The product in the integrand can be expanded and written as
$G=[1-w'\sum_k \exp(-i 2\pi k/K) ...+(-w')^K\exp(-i2\pi\sum_{k=1}^K k)/K ]^{-2/K}.$
For a regular polygon, only the first and last terms within the square brackets remain.
The terms in the middle sum to zero.  We get
$G=[1+(-w')^K\exp(-i \pi (K+1)) ]^{-2/K}=[1-(w')^K ]^{-2/K}.$
We compared our result in Figure \ref{val} for the square with K=4 when the conformal mapping can be obtained analytically (SC) by numerically integrating Eq. (\ref{scz}) with results using computational conformal mapping (CEM) recapitulated in Appendix \ref{A:CEM}  and
found good agreement.
The derivative $dz/dw=[1-(w)^K ]^{-2/K}$ with the SC transformation is also found to agree with the result with the CEM method to within 5 per cent.
\subsubsection{Boundary electric ﬁelds}
\begin{figure}[tbph]
\vspace*{0pt} \centerline{\includegraphics[angle=0,width=8cm]{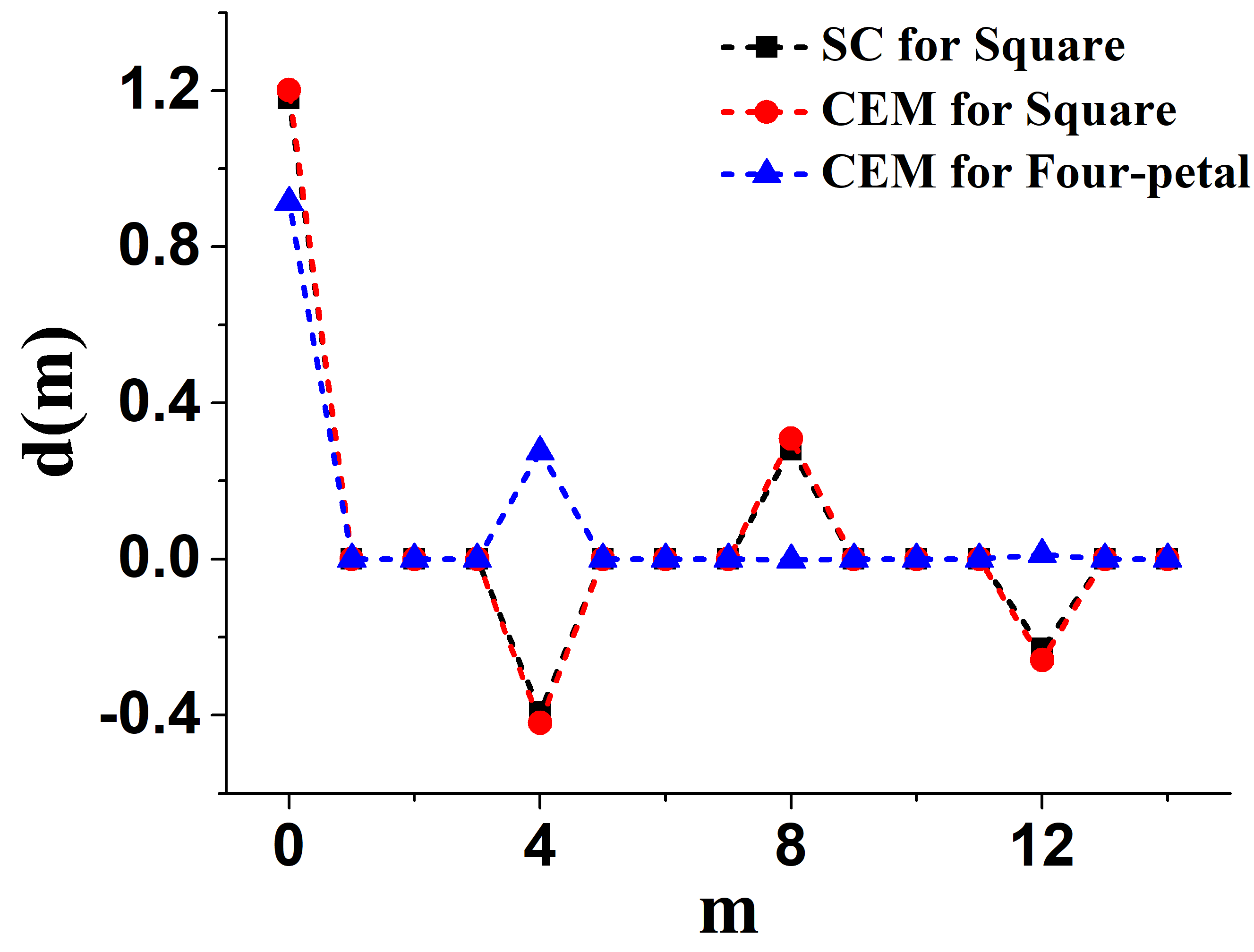}}
\vspace*{0pt}
\caption{{\bf Fourier components $\bm{d(m)}$.} Profiles $d(m)$ of the circular basis vectors projected onto the normal of the square plate (black dot-dashed line: SC transformation; red dot-dashed line: CEM method) and four-petal plate (blue dot-dashed line: CEM method).}
\label{fig:d(m)}
\end{figure}
The electric field's behavior at the edges and corners of finite surfaces has been elucidated\cite{chui2016scattering} previously. By substituting the boundary normal vector $\mathbf{e}_{R}^{z}=\left(e^{i \phi_{w}} p_{+} \mathbf{e}_{-}+e^{-i \phi_{w}} p_{-} \mathbf{e}_{+}\right) /(2|p|)$ into Eq. \eqref{eq:Bint} and simplifying it, we get expressions for the boundary electric field contributions $B_{M(N)}$.
\begin{equation}
\begin{aligned}
B_{M}(m, l, \kappa) & =\frac{R m J_{m}(k_{\kappa} R) d(m-l)}{k R C_{M m}},   \\
B_{N}(m, l, \kappa) & =\frac{R J_{m}^{\prime}(k_{\kappa} R) d(m-l)}{C_{N m}},
\end{aligned}
\label{eq:BMN}
\end{equation}
where $d(m)=\int d \phi_{w} \exp \left(im \phi_{w}\right) /|p| /(2 \pi)$. It is the integrable Jacobian factor $|p|=|d w / d z|$ in the denominator of $d(m)$ that causes the enhancement of the ﬁeld at the corners. By using the conformal transformation, this ``singularity'' is automatically incorporated.

 Figure \ref{fig:d(m)} shows the profiles of $d(m)$ for both shapes. For the square, the Jacobian factor $|p|$ is retrievable from the SC transformation and the CEM method, both of which yield $d(m)$ results that agree with each other, as shown in Figure \ref{fig:d(m)} with the black and red dot-dashed lines, respectively. The results validate the effectiveness of the CEM algorithm. For the four-petal shape, the Jacobian factor is challenging to obtain through the SC transformation, yet it can be successfully computed using the CEM method (as shown by the blue dot-dashed line in Figure \ref{fig:d(m)}). This highlights the advantage of our newly proposed method in handling a variety of geometric shapes beyond just polygons.

The non-vanishing of $d(m)$ for m's that are multiples of four for both shapes indicates a congruence in their fourfold rotation symmetry, which are crucial for understanding the electromagnetic behavior of these shapes.

\subsubsection{Green's function matrix (Circuit Parameters)}
\begin{figure}[tbph]
\vspace*{0pt} \centerline{\includegraphics[angle=0,width=11cm]{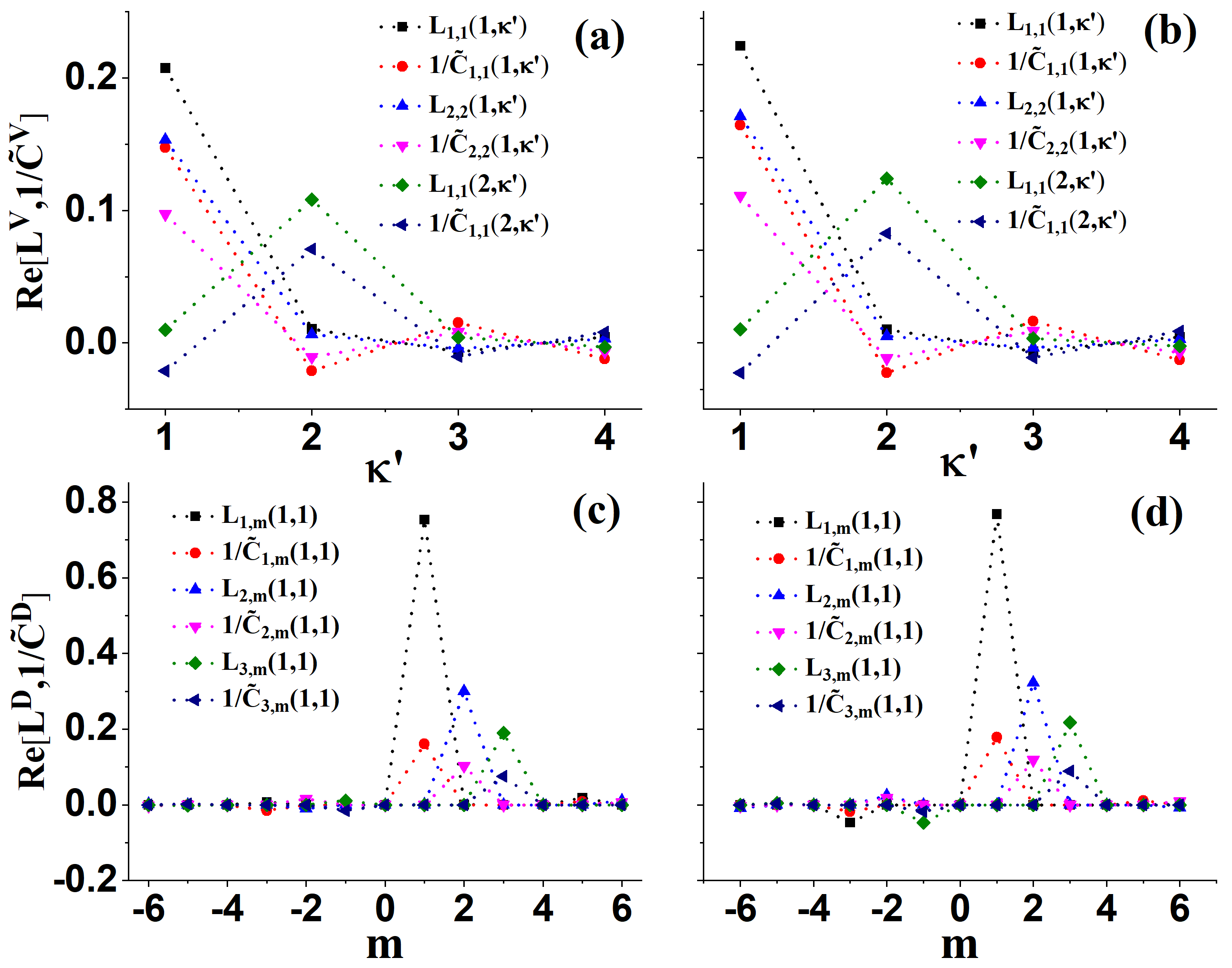}}
\vspace*{0pt}
\caption{{\bf Circuit parameters of the square plate under two conformal mapping methods.} (a) and (b) display $L_{X m, X m}\left(\kappa,\kappa^{\prime}\right)$ and $\tilde{C}_{N m, N m}^{-1}\left(\kappa,\kappa^{\prime}\right)$ for the V-type boundary condition; (c) and (d) illustrate $L_{X n, X m}\left(\kappa,\kappa\right)$ and $\tilde{C}_{N n, N m}^{-1}\left(\kappa,\kappa\right)$ for the D-type boundary condition. The conformal mapping methods applied in (a) and (c) are based on the SC transformation; those in (b) and (d), on the CEM. The units of the matrices $\mathbf{L}$ and $\mathbf{\tilde{C}}^{-1}$ are $2\pi\mu_{0}\epsilon^{-1}_{0}$ and $2\pi\epsilon^{-1}_{0}$ respectively. }
\label{fig:Zmn}
\end{figure}
The “impedance” matrix $\mathbf{Z}^{0}$ in Eq. \eqref{eq:Cir} can be expressed as the sum of an inductive $\mathbf{L}$ and a capacitive $\mathbf{C}^{-1}$ term: $\mathbf{Z}^{0}=i \omega \mathbf{L}+i /(\omega \mathbf{C})$. The elements of these matrices are just the representation of the Green's function $ G_{0}\left(\mathbf{r}, \mathbf{r}^{\prime}\right)=$ $\exp \left(i k_{0}\left|\mathbf{r}-\mathbf{r}^{\prime}\right|\right) /\left(4\pi\epsilon_0\left|\mathbf{r}-\mathbf{r}^{\prime}\right|\right)$ in our basis functions $\mathbf{cX}, \mathbf{cY}=\mathbf{c M}, \mathbf{c N}$:
\begin{align}
L_{X n, Y m}\left(\kappa,\kappa^{\prime}\right) &=\mu_{0}
\iint d \mathbf{r} d \mathbf{r}^{\prime} \mathbf{cX}^{*}_{n}\left[k_{\kappa} w(z)\right] \cdot \mathbf{cY}_{m}\left[k_{\kappa'} w^{\prime}(z^{\prime})\right] G_{0}\left(\mathbf{r}, \mathbf{r}^{\prime}\right),\label{eq:L}\\
C_{X n, Y m}^{-1}\left(\kappa,\kappa^{\prime}\right) &=
 \iint d \mathbf{r} d \mathbf{r}^{\prime} \mathbf{cX}^{*}_{n}\left[k_{\kappa} w(z)\right] \cdot \nabla \nabla^{\prime} \cdot \mathbf{cY}_{m}\left[k_{\kappa'} w^{\prime}(z^{\prime})\right] G_{0}\left(\mathbf{r}, \mathbf{r}^{\prime}\right),
 \label{eq:C-1}
\end{align}
here, the subscripts $X,Y=M,N$; the position vector $\mathbf{r}=x\mathbf{e}_{x}+y\mathbf{e}_{y}$ lies on the surface, and the corresponding variable $z=\mathbf{r}\cdot \mathbf{e}_{+}$. Thus $G_{0}\left(\mathbf{r}, \mathbf{r}^{\prime}\right)=G_{0}\left(z, z^{\prime}\right)$. 
Combining conformal maps and Eq. \eqref{eq:cMcN}, we convert the integration \eqref{eq:L} and \eqref{eq:C-1} over the finite surface to that over a disk. We find that the inductances with
respect to the basis functions $\mathbf{cM}, \mathbf{cN}$ contain contributions from the right and left circularly polarized components and thus can be written as
\begin{equation}
L_{X n, Y m}\left(\kappa,\kappa^{\prime}\right)  =L_{n-1, m-1}\left(\kappa,\kappa^{\prime}\right)+\left(2\delta_{X,Y} - 1\right)L_{n+1, m+1}\left(\kappa,\kappa^{\prime}\right), X,Y=M, N
\nonumber  
\end{equation}
where
\begin{equation}
L_{n, m}\left(\kappa,\kappa^{\prime}\right)  =\mu_{0} \iint d \boldsymbol{w} d \boldsymbol{w}^{\prime} f_{n}^{*}(k_{\kappa} w) G_{0}\left(z(w), z^{\prime}(w^{\prime})\right) f_{m}\left(k_{\kappa'} w^{\prime}\right) /\left[p_{-}(w) p_{+}\left(w^{\prime}\right)\right].
\label{eq:Lmn}
\end{equation}
The only nonzero components of the capacitance term is given by
\begin{equation}
C_{N n, N m}^{-1}\left(\kappa,\kappa^{\prime}\right)  =-k_{\kappa} k_{\kappa'}
\iint d \boldsymbol{w} d \boldsymbol{w}^{\prime} f_{n}^{*}(k_{\kappa} w) G_{0}\left(z(w), z^{\prime}(w^{\prime})\right) f_{m}\left(k_{\kappa'} w^{\prime}\right).
\label{eq:C-1mn}
\end{equation}

The circuit parameters with the analytic conformal mapping can be calculated in a straightforward way as in our previous work with a well developed algoraithm that takes care of the integrable singularity of the Green's function $G_0(|\br-\br'|)$ for $\br=\br'$.
To calculate the circuit parameters for triangular meshes generated with the CEM method for various shapes, we approximate the integral operations over a disk as discrete summations. This approach facilitates the embedding of conformal transformations from arbitrary shapes to a disk. We perform a triangulation mesh for each shape with N triangular elements in a certain order, as illustrated in Figure \ref{fig:diskmaps}. For the $i$th triangular element within the disk, we define its area as $\Delta \sigma_{i}$ and the position point of its centroid as $w_i$. Using this approach, we can express the inductance and capacitance-related integrals as follows respectively
\begin{align}
L_{n, m} \left(\kappa,\kappa^{\prime}\right) &= \mu_{0}\sum_{i}^{N} \sum_{j}^{N} \frac{f_{n}^{*}\left(k_{\kappa} w_{i}\right) G_{0}\left(z_{i}, z_{j}\right) f_{m}\left(k_{\kappa'} w_{j}\right) \Delta \sigma_{i} \Delta \sigma_{j}}{p_{-}\left(w_{i}\right) p_{+}\left(w_{j}\right)}, \label{eq:S-Lmn}\\
C_{N n, N m}^{-1}\left(\kappa,\kappa^{\prime}\right) & =-k_{\kappa} k_{\kappa'} \sum_{i}^{N} \sum_{j}^{N} f_{n}^{*}\left(k_{\kappa} w_{i}\right) G_{0}\left(z_{i}, z_{j}\right) f_{m}\left(k_{\kappa'} w_{j}\right) \Delta \sigma_{i} \Delta \sigma_{j}.
\label{eq:S-C-1mn}
   \end{align}
The CEM method is recapitulated in Appendix \ref{A:CEM}. Briefly, we construct a discrete conformal energy function for a Delaunay triangular
mesh. By solving the minimization problem of this function, we obtain the desired conformal mapping between $z_i$ and $w_i$ and the corresponding derivatives $p_{\pm}$.  For the $i = j$ terms in the summations in Eqs \eqref{eq:S-Lmn} and \eqref{eq:S-C-1mn}, the singularity $G_{0}\left(z_{i}, z_{i}\right)$ requires special attention. We  approximated the field by a constant value and performed the integration of the Green's function over the triangle. The detailed procedure is provided in Appendix \ref{B:Griri}.
\begin{figure}[tbph]
\vspace*{0pt} \centerline{\includegraphics[angle=0,width=11cm]{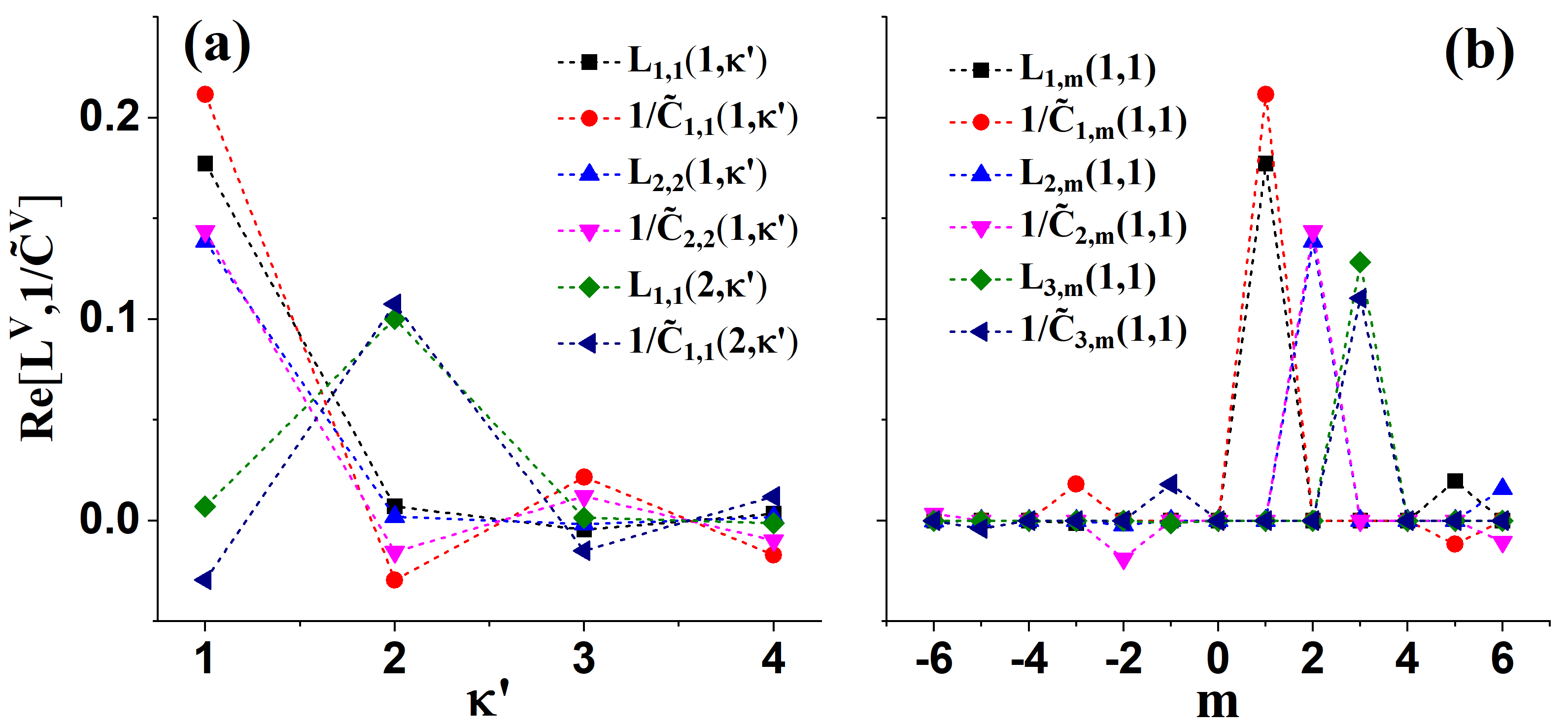}}
\vspace*{0pt}
\caption{{\bf Circuit parameters for the four-petal plate with the CEM-based conformal mapping.} (a) displays $L_{X m, X m}\left(\kappa,\kappa^{\prime}\right)$ and $\tilde{C}_{N m, N m}^{-1}\left(\kappa,\kappa^{\prime}\right)$ for the V-type boundary condition; (b) illustrates $L_{X n, X m}\left(\kappa,\kappa\right)$ and $\tilde{C}_{N n, N m}^{-1}\left(\kappa,\kappa\right)$ for the V-type boundary condition. The units of the matrices $\mathbf{L}$ and $\mathbf{\tilde{C}}^{-1}$ are $2\pi\mu_{0}\epsilon^{-1}_{0}$ and $2\pi\epsilon^{-1}_{0}$ respectively.}
\label{fig:ZmnNS}
\end{figure}

We have numerically calculated the circuit parameters for Eqs. \eqref{eq:S-Lmn} and \eqref{eq:S-C-1mn} by efficiently employing parallel computing. Different elements can be calculated on different processors, no interprocess communication is required. 
Figure \ref{fig:Zmn} illustrates the $\kappa$ and $m$ dependence of the circuit matrix elements of a square metallic plate.  Figures \ref{fig:Zmn}(a) and \ref{fig:Zmn}(c) present the V-type and D-type circuit matrices $L_{X n, X m}\left(\kappa,\kappa^{\prime}\right)$ and $\tilde{C}_{N n, N m}^{-1}\left(\kappa,\kappa^{\prime}\right)=-C_{N n, N m}^{-1}\left(\kappa,\kappa^{\prime}\right)/(k_{\kappa} k_{\kappa'})$ obtained by numerically integrating Eqs \eqref{eq:Lmn} and \eqref{eq:C-1mn} 
with the SC transformation. Figures \ref{fig:Zmn}(b) and \ref{fig:Zmn}(d) display the corresponding circuit matrices $L_{X n, X m}\left(\kappa,\kappa^{\prime}\right)$ and 
$\tilde{C}_{N n, N m}^{-1}\left(\kappa,\kappa^{\prime}\right)$ from equations \eqref{eq:S-Lmn} and \eqref{eq:S-C-1mn} with the CEM algorithm. These figures not only confirm that the impedance matrix are indeed nearly diagonal, but also reveal that the inductance and capacitance matrix, derived from the basis functions generated by the two mappings, agree with each other to within 5\%. Furthermore, the circuit elements also decreases as the angular momentum index $m$ is increased, thus only a few $m$ terms are required for an accurate result. $C_{N n, N m}^{-1}\left(\kappa,\kappa^{\prime}\right)$ increases rapidly as $k_{\kappa},k_{\kappa'}$ is increased. This makes the problem of inverting the impedance matrix rapidly convergent and is one of the simplifying feature of the present approach.

The CEM algorithm not merely is an alternative to the SC transformation for polygonal geometries but also facilitates efficient computations for more general shapes. We have applied the CEM algorithm to calculate the circuit parameter matrices of a four-petal metallic plate, which is challenging to handle using the SC transformation. Figures \ref{fig:ZmnNS}(a) and \ref{fig:ZmnNS}(b) show the calculated results of the V-type inductance $L_{X n, X m}\left(\kappa,\kappa^{\prime}\right)$ and capacitance $\tilde{C}_{N n, N m}^{-1}\left(\kappa,\kappa^{\prime}\right)$, indicating that both matrices are nearly diagonalized and exhibit a decrease with increasing index $\kappa$ and $m$.

\subsubsection{Numerical Results for the Resonance Behavior}

 The resonance behavior of both square and four-petal metallic plates with a dimensionless side length of $a=1.81$, corresponding to a disk of unit radius, has been investigated.
The low lying resonances are determined by the sign change in the real part of eigenvalues of $\mathbf{K}$ in Eq. \eqref{eq:Es&K} over a range of frequencies. The imaginary part of the eigenvalues, mainly due to weak impact of intrinsic and radiative resistances, cause less than 1\% change in resonance frequencies\cite{chui2016scattering}. We have included the lowest four $k$ values and set the range of angular momentum $m$ from -6 to +6 to ensure the convergence of the results. In the results, we have identified the eigenfunction of the matrix ${\bf K}$, the``bare'' boundary field, for various resonance modes with eigenvalues close to zero, some of which are degenerate.
\begin{figure}[tbph]
\vspace*{0pt} \centerline{\includegraphics[angle=0,width=11cm]{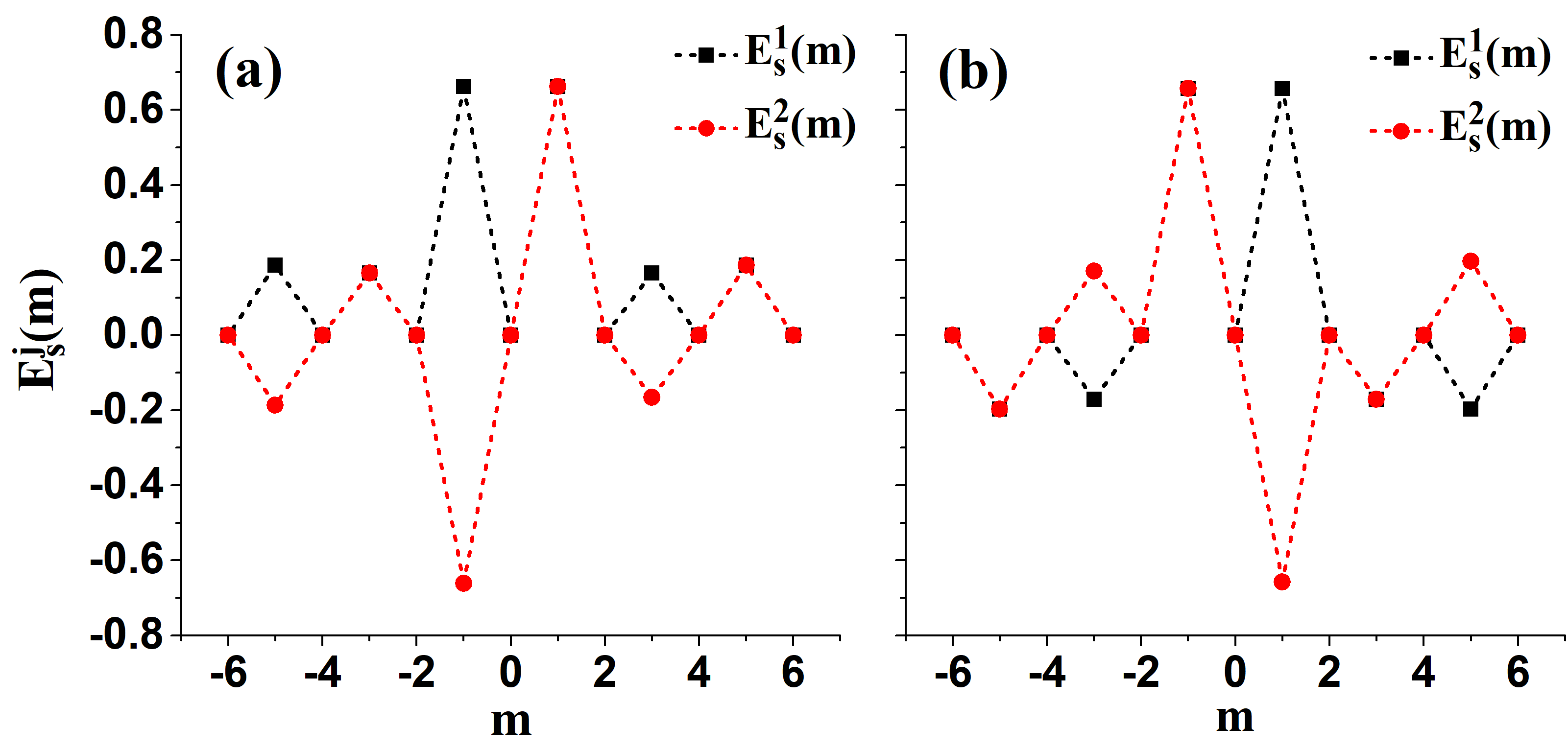}}
\vspace*{0pt}
\caption{{\bf Angular momentum components of ``bare'' boundary field.} Eigenfunctions of the matrix ${\bf K}$, the ``bare'' boundary field, for the lowest two resonance modes with eigenvalues near zero in (a) the square plate and (b) the four-petal plate. For the case when there are degenerate resonances.}
\label{fig:Ejm}
\end{figure}
For the V(D)-type boundary condition, from Eq. \eqref{eq:BMN} $B_{M(N)}=0$, $\mathbf{K}$ involves only terms with the $N(M)$ component of the basis functions
\begin{equation}
 \mathbf{K}^{V(D)}=\mathbf{B}_{N(M)} \mathbf{Z}^{-1} \mathbf{B}_{N(M)}.
\label{eq:K}
\end{equation}

According to the resonance condition given by Eq. \eqref{eq:detK}, the lowest resonances of the two plates come from the D-type boundary condition and are both doubly degenerate. The lowest normalized resonance frequencies for the square, four-petal plates are given by $\omega_{r}a/c\approx2.11, 2.57$ respectively. The information on the degeneracy is difficult to extract from conventional numerical approaches but is trivial in our approach. The boundary field components $E_{S}(m)$ of the two degenerate resonances for each shape are shown in Figure \ref{fig:Ejm}(a) and \ref{fig:Ejm}(b), respectively. It can be observed that both shapes, possessing the same fourfold rotational symmetry, exhibit identical properties, with $E_{S}^{1}(m)=E_{S}^{1}(-m)$ and $E_{S}^{2}(m)=-E_{S}^{2}(-m)$. These two degenerate modes are mainly dominated by $|m|=1$ states and correspond to linear combinations of ($m=\pm1,\pm3,\pm5, \ldots$).
\begin{figure}[tbph]
\vspace*{0pt} \centerline{\includegraphics[angle=0,width=11cm]{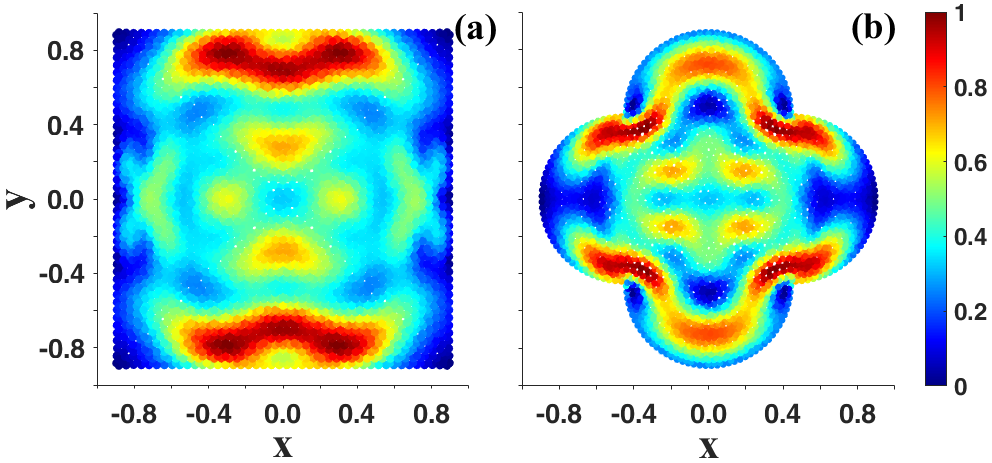}}
\vspace*{0pt}
\caption{{\bf The normalized lowest resonance field magnitudes.} Magnitude distributions of the lowest resonance current fields $|\mathbf{j}|$ in (a) the square plate and (b) the four-petal plate. Both plates share an identical side length $a=1.81$. Field magnitudes are normalized to their respective maximum values (a.u.).}
\label{fig:Resf}
\end{figure}
\begin{figure}[tbph]
\vspace*{0pt} \centerline{\includegraphics[angle=0,width=11cm]{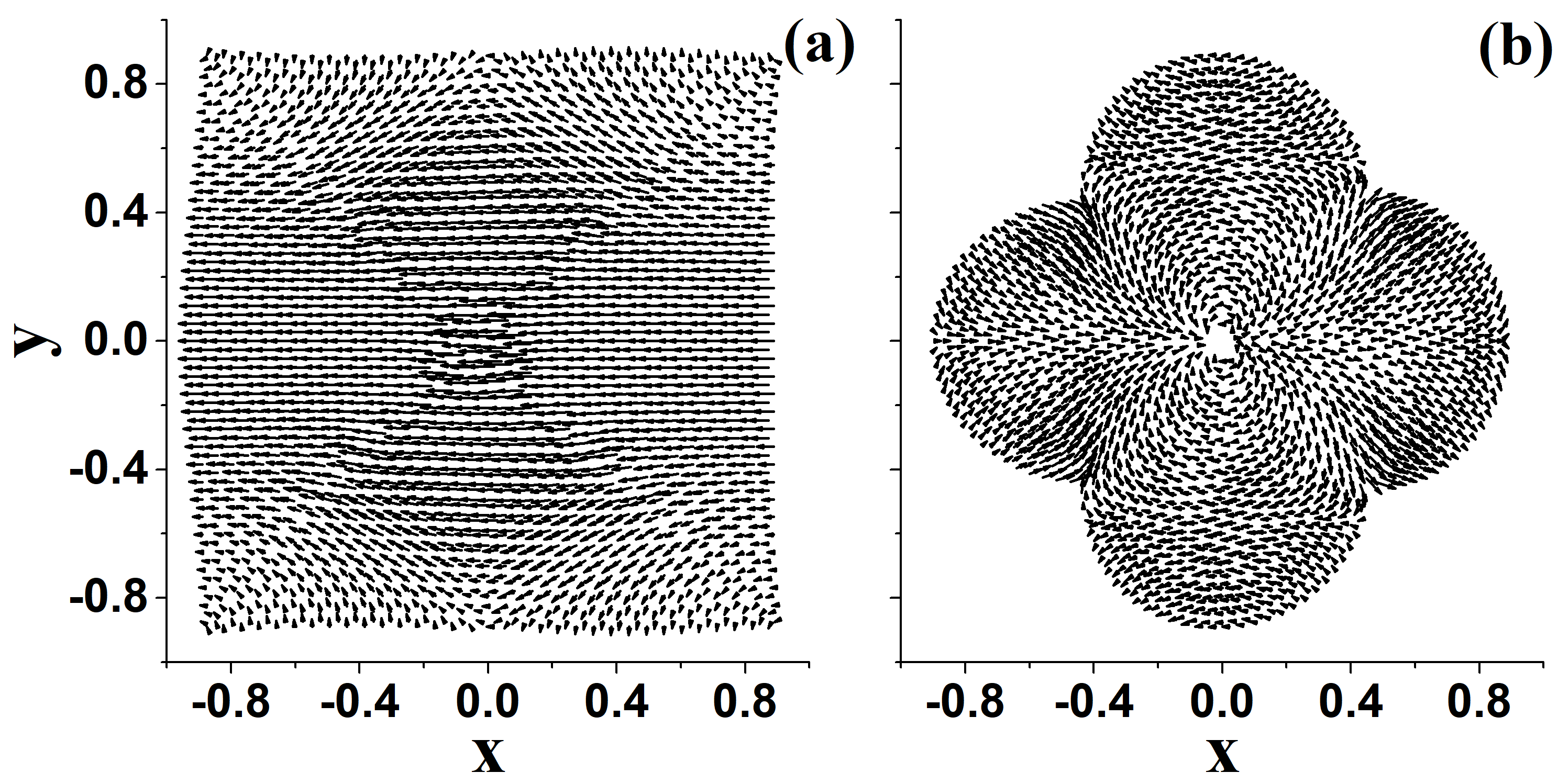}}
\vspace*{0pt}
\caption{{\bf Real parts of the lowest resonance current components.} (a) Real component of the resonance current $\mathbf{j}_{M,m=1,\kappa=1}$ for the square plate, exhibiting ``fractional'' vorticity localized near the corners. (b) Real component of $\mathbf{j}_{N,m=3,\kappa=1}$ for the four-petal plate, showing current divergence(convergence) close to the corners. Both plates share an identical side length $a=1.81$. Field magnitudes are normalized for visualization (a.u.).}
\label{fig:VecF}
\end{figure}
The $E_{S}(m)$ corresponds to a rich pattern of electric currents inside the shapes. Based on Eq. \eqref{eq:Cir}, when the external ﬁeld $\mathbf{E}_{\text {ext }}$  becomes very small, the resonance current $\mathbf{j}$ can be expressed as
\begin{equation}
\mathbf{j}=\mathbf{Z}^{-1} \mathbf{E}_{s}.
 \label{eq:jr}
\end{equation}
The lowest resonance current field magnitudes $\mathbf{|j|}$ for the square and the four-petal, calculated by Eq. \eqref{eq:jr}, are displayed in Figure \ref{fig:Resf}(a) and \ref{fig:Resf}(b) respectively. The field magnitudes exhibit maximal enhancement bilaterally near their geometric edges. 

We can even analyze the distribution characteristics of each component constituting the resonant current. From Eq. \eqref{eq:jcMN}, an $m$ component current field $\mathbf{j}_m$ can be expressed as a linear combination of divergence-free $\mathbf{cM}_m$ and curl-free $\mathbf{cN}_m$, with corresponding local curl and divergence given by
\begin{equation}
\begin{aligned}
\nabla \times \mathbf{cM}_m=i k_{\kappa} \mathbf{e}_{z} |p|^{2} f_{m},   \\
\nabla \cdot \mathbf{cN}_m=-k_{\kappa} |p|^{2} f_{m},
\end{aligned}
\label{eq:CuDi}
\end{equation}
here the unit vector $\mathbf{e}_{z}=\mathbf{e}_{x}\times\mathbf{e}_{y}$ perpendicular to the x-y plane. This allows systematic extraction of the modes for each $m$ component. Figure \ref{fig:VecF}(a) displays the fundamental ($m,\kappa=1$) $\mathbf{cM}$-component resonance current $\mathbf{j}_{M,m=1,\kappa=1}$ in the square, exhibiting fractional vortex patterns localized near the corners. Conversely, Figure \ref{fig:VecF}(b) shows the corresponding ($m=3,\kappa=1$) $\mathbf{cN}$-mode $\mathbf{j}_{N,m=3,\kappa=1}$, which manifests current divergence/convergence at junctions of the four-petal structure. 
Eqs. \eqref{eq:CuDi} indicate that the local vorticity(divergence) of the current field in the square(four-petal) arises from two contributions: a ``global'' contribution from the factor $f_{m}$ and a ``local'' contribution from the Jacobian $|p|$. Crucially, this shape-dependent $|p|$ contribution vanishes in rotationally symmetric disks where $|p|=1$, highlighting geometric singularities in tailoring electromagnetic fields.  
\subsubsection{Results from COMSOL methods}
\begin{figure}[tbph]
\vspace*{0pt} \centerline{\includegraphics[angle=0,width=11cm]{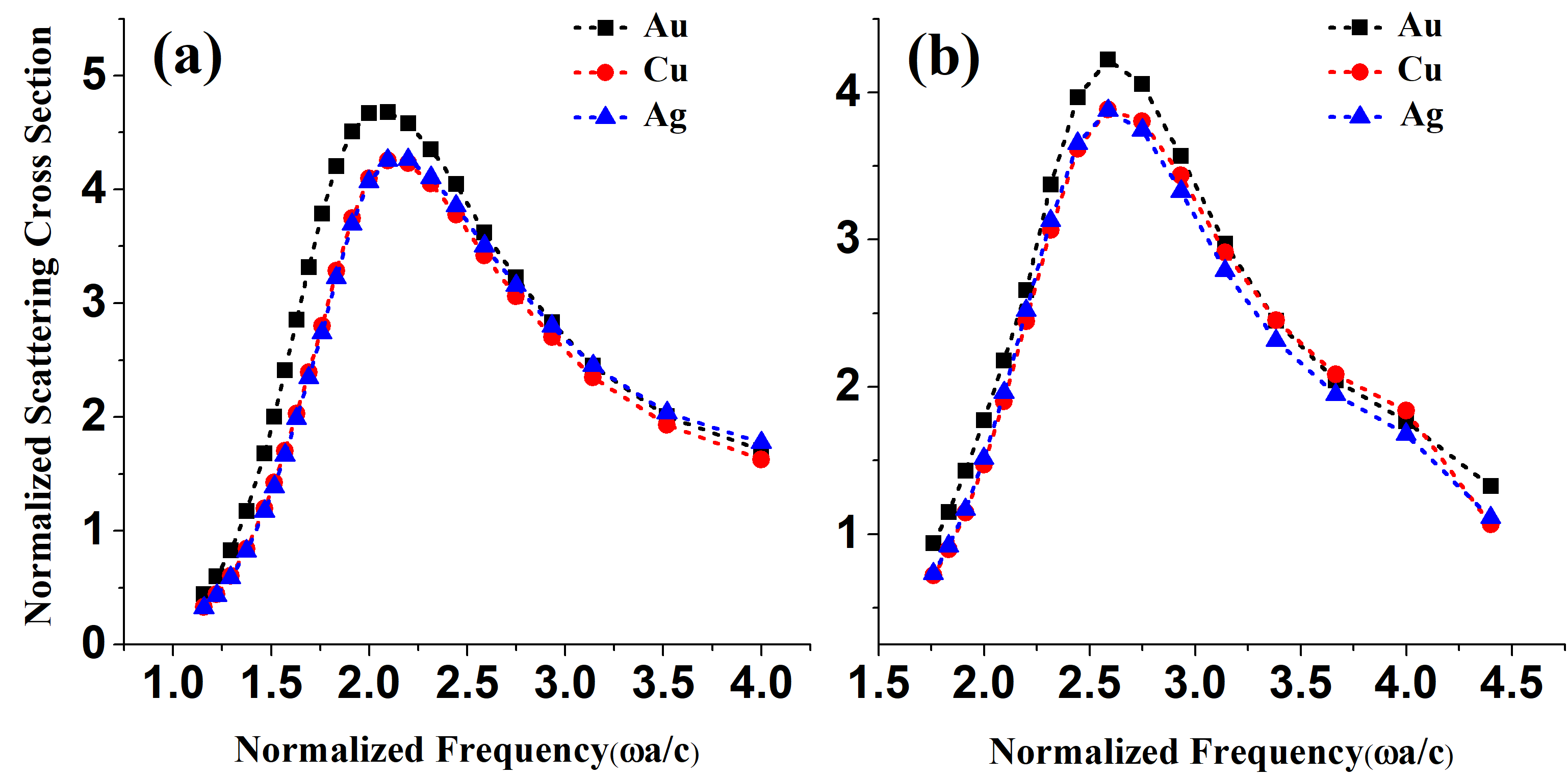}}
\vspace*{0pt}
\caption{{\bf Finite element simulations for the two metallic plates.} Normalized scattering cross section obtained by COMSOL for (a) the square plate and (b) the four-petal plate.}
\label{fig:NSCS}
\end{figure}

We have performed finite element simulations using COMSOL Multiphysics®6.2 to compute the normalized scattering cross section (NSCS) for a square plate and a four-petal plate, each with a side length of 700 nm and a thickness of 21 nm, fabricated from Au(gold), Cu(copper), and Ag(silver). Figures \ref{fig:NSCS}(a) and \ref{fig:NSCS}(b) show the frequency-dependent NSCS profiles for the square plate and the four-petal plate, respectively. They exhibit peaks at the normalized frequencies $\omega_{r}a/c\approx2.10$ for the square and $\omega_{r}a/c\approx2.59$ for the four-petal, close to the lowest resonance frequency obtained by our method for the each plate. Notably, the computational efficiency of our approach achieves a breakthrough, requiring only 15 ms per frequency – three orders of magnitude faster than conventional finite element methods (FEM) requiring 60 s per frequency. 

\section{Conclusions}

In this paper we demonstrated with two examples a {\bf general} way to solve the electromagnetic scattering problem that is three orders of magnitude faster than current finite element methods. Using computational conformal geometry (via the CEM algorithm)
we show how previous approaches for special cases can be applied to general surfaces. This opens the door to  important practical real time applications where the speed of the computation is a major obstacle.
With the CEM algorithm we generate orthonormal basis functions for general surfaces via conformal mapping to simple surfaces whose orthonormal basis function is known. The electromagnetic field and the Green's function are expressed in this basis. The boundary condition is implemented with the introduction of self-consistently determined auxillary boundary electric fields. The integral form of Maxwell's equation is solved.

Our approach reveals low-energy degenerate resonance modes that are elusive to traditional methods, thereby enriching the understanding of EM behavior in complex geometries. It is more accurate for surfaces with sharp corners where the singularity is built in with the basis functions constructed with the conformal mapping.

\appendix  
\setcounter{equation}{0}  
\setcounter{figure}{0}    
\section{Conformal Maps by CEM}\label{A:CEM}
\renewcommand{\theequation}{A.\arabic{equation}}
\renewcommand{\thefigure}{A.\arabic{figure}}  
In this section, we briefly recapitulate the conformal map and its computation by conformal energy minimization. Please refer to \cite{Hutc91,YCWW21} for details.
Let $\mathcal{M}, \mathcal{N} \subset \mathbb{R}^2$ be simply connected domains. Then the map $f: \mathcal{M} \to \mathcal{N}$ is called conformal if $f(v^1,v^2) = (x(v^1,v^2),y(v^1,v^2))$ satisfies the Cauchy-Riemann equations
\begin{align}
    \frac{\partial x}{\partial v^1} = \frac{\partial y}{\partial v^2}, \quad \frac{\partial x}{\partial v^2} = -\frac{\partial y}{\partial v^1}. \label{eq:CReqn}
\end{align}
From the Cauchy-Riemann equations in \eqref{eq:CReqn}, the Jacobian matrix of $f$ becomes
\begin{align}
    J_f = \begin{bmatrix}
        \frac{\partial x}{\partial v^1} & \frac{\partial x}{\partial v^2} \\
        \frac{\partial y}{\partial v^1} & \frac{\partial y}{\partial v^2}
    \end{bmatrix}
    =
    \begin{bmatrix}
        a & b \\
        -b & a
    \end{bmatrix}, \text{ with } a = \frac{\partial x}{\partial v^1}, b = \frac{\partial x}{\partial v^2}, \nonumber
\end{align}
which is the composite of a scalar and an orthogonal matrix.
To compute the conformal map, we consider the least squares of the Cauchy-Riemann equations
\begin{align}
    E_C(f) :=& \frac{1}{2} \int_\mathcal{M} \left[\left( \frac{\partial x}{\partial v^1} - \frac{\partial y}{\partial v^2} \right)^2 + \left( \frac{\partial x}{\partial v^2} + \frac{\partial y}{\partial v^1} \right)^2 \right] ds \label{eq:CE-def}\\
    = & \frac{1}{2} \int_\mathcal{M} \left[ \left(\frac{\partial x}{\partial v^1}\right)^2 +
    \left(\frac{\partial y}{\partial v^2}\right)^2 +
    \left(\frac{\partial x}{\partial v^2}\right)^2 +
    \left(\frac{\partial y}{\partial v^1}\right)^2 \right] ds  \nonumber \\
    & - \int_\mathcal{M} \left[\frac{\partial x}{\partial v^1} \frac{\partial y}{\partial v^2} - \frac{\partial x}{\partial v^2}  \frac{\partial y}{\partial v^1} \right] ds \nonumber \\
    =& \frac{1}{2} \int_\mathcal{M} \|J_f\|_F^2 ds - \int_\mathcal{M} \det(J_f) ds = E_D(f) - |\mathcal{N}|, \nonumber
\end{align}
where $E_C(f)$ is the conformal energy functional, $E_D(f)$ is the Dirichlet energy functional and $|\mathcal{N}|$ represents the area of the target domain $\mathcal{N} = f(\mathcal{M})$.
\eqref{eq:CE-def} illustrates the fact that the conformal energy functional $E_C(f)$ is nonnegative and equals to $0$ if and only if $f$ is a conformal map. Therefore, we can minimize the conformal energy functional to compute the conformal map.

In the discrete version,
the domain $\mathcal{M}$ is the geometric figure of interest. It is characterized by a Delaunay triangular
mesh with $n$ vertices $v_{s} \in \mathbb{R}^2$.
Let $\mathcal{V}(\mathcal{M}), \mathcal{F}(\mathcal{M}), \mathcal{E}(\mathcal{M})$ denote the sets of vertices, faces, and edges of $\mathcal{M}$, respectively,
\begin{align}
\mathcal{V}(\mathcal{M})=\left\{v_{s}\right\}_{s=1}^{n},\quad
\mathcal{F}(\mathcal{M})=\left\{\tau_{ijk} := \left[v_{i}, v_{j}, v_{k}\right]\right\}, \quad
\mathcal{E}(\mathcal{M})=\left\{\left[v_{i}, v_{j}\right] | \left[v_{i}, v_{j}, v_{k}\right] \in \mathcal{F}(\mathcal{M})\right\}. \nonumber
\end{align}
The map $f$ is considered to be a piecewise linear map, which is linear on each triangle $\tau \in \mathcal{F}(\mathcal{M})$. Then the discrete conformal energy can be written as
\begin{align}
    E_C(f) = \frac{1}{2} x^\mathrm{T} L x + \frac{1}{2} y^\mathrm{T} L y - x^\mathrm{T} D y \label{eq:EC_discrete}
\end{align}
with
\begin{align}
    \big[ L \big]_{ij} = \begin{cases}
        -w_{ij} := -\frac{1}{2}(\cot\alpha_{ij} + \cot\alpha_{ji}), & [v_i,v_j] \in \mathcal{E}(\mathcal{M}) \\
        \sum_{j = 1}^n w_{ij}, & i = j\\
        0,& \mathrm{otherwise}
    \end{cases}, \
    \big[ D \big]_{ij} = \begin{cases}
        \frac{1}{2}, & [v_i,v_j] \in \mathcal{E}(\partial \mathcal{M}), v_i \to v_j \text{ is anticlockwise} \\
        -\frac{1}{2}, & [v_i,v_j] \in \mathcal{E}(\partial \mathcal{M}), v_i \to v_j \text{ is clockwise}\\
        0,& \mathrm{otherwise}
    \end{cases}, \nonumber
\end{align}
where $\alpha_{ij}$ and $\alpha_{ji}$ are the opposite angles with respect to the edge $[v_i,v_j]$.
Partition the vertices $\mathcal{V}(\mathcal{M})$ into boundary and interior parts and let $\mathtt{B} = \{i | v_i \in \mathcal{V}(\partial \mathcal{M})\}$ and $\mathtt{I} = \{i | v_i \in \mathcal{V}( \mathcal{M} \setminus\partial \mathcal{M})\}$. Then $x$, $y$, $L$ and $D$ can also be partitioned accordingly,
\begin{align}
    x = \begin{bmatrix}
        x_{\mathtt{B}} \\
        x_{\mathtt{I}} \\
    \end{bmatrix},\quad
    y = \begin{bmatrix}
        y_{\mathtt{B}} \\
        y_{\mathtt{I}} \\
    \end{bmatrix},\quad
    L = \begin{bmatrix}
        L_{\mathtt{B}\mathtt{B}} & L_{\mathtt{B}\mathtt{I}} \\
        L_{\mathtt{I}\mathtt{B}} & L_{\mathtt{I}\mathtt{I}} \\
    \end{bmatrix},\quad
    D = \begin{bmatrix}
        D_{\mathtt{B}\mathtt{B}} & \mathbf{0} \\
        \mathbf{0} & \mathbf{0} \\
    \end{bmatrix}.  \label{eq:LD_partition}
\end{align}

With regard to our goal, that is, computing the disk conformal map, we impose the circular constraint $\{x_i^2 + y_i^2 = 1, i \in \mathtt{B}\}$ on the conformal energy minimization.
Naturally, the boundary vertices $(x_\mathtt{B},y_\mathtt{B})$ can be represented by the polar coordinates, and their gradients are also represented by $(x_\mathtt{B}, y_\mathtt{B})$, i.e.,
\begin{align} \label{eq:polar}
\begin{cases}
    x_i = \cos\theta_i\\
    y_i = \sin\theta_i
\end{cases}, \quad
\begin{cases}
    \nicefrac{dx_i}{d\theta_i} = -y_i\\
    \nicefrac{dy_i}{d\theta_i} = x_i
\end{cases}
    v_i \in \mathcal{V}(\partial \mathcal{M}).
\end{align}
From \eqref{eq:EC_discrete} and \eqref{eq:LD_partition}, the optimal interior vertices $(x_\mathtt{I},y_\mathtt{I})$ should satisfy
\begin{align}
    \nabla_{[x_\mathtt{I},y_\mathtt{I}]} E_C(f) = L_{\mathtt{I}\mathtt{I}} [x_{\mathtt{I}},
 y_{\mathtt{I}}] + L_{\mathtt{I}\mathtt{B}} [x_{\mathtt{B}}, y_{\mathtt{B}}] = 0. \label{eq:KKT-int}
\end{align}
Subsequently, by the equation in \eqref{eq:KKT-int} and the polar coordinate representation in \eqref{eq:polar}, the conformal energy can be formulated with respect to the polar angles $\theta$ only, i.e.,
\begin{align}
    E_C(\theta) = \frac{1}{2} x^\mathrm{T}_{\mathtt{B}} S x_{\mathtt{B}} + \frac{1}{2} y^\mathrm{T}_{\mathtt{B}} S y_{\mathtt{B}} - x^\mathrm{T}_{\mathtt{B}} D_{\mathtt{B}\mathtt{B}} y_{\mathtt{B}}, \label{eq:EC_discrete2}
\end{align}
where $S = L_{\mathtt{B}\mathtt{B}} - L_{\mathtt{B}\mathtt{I}}^\mathrm{T} L_{\mathtt{I}\mathtt{I}}^{-1} L_{\mathtt{I}\mathtt{B}}$ is the Schur complement of $L$.
The gradient of the conformal energy in \eqref{eq:EC_discrete2} with respect to $\theta$ becomes
\begin{align}
    \nabla_{\theta} E_C(\theta)
    =& -\diag(y_\mathtt{B}) S x_{\mathtt{B}}
    + \diag(x_\mathtt{B}) S y_{\mathtt{B}}
    - \diag(y_\mathtt{B}) D_{\mathtt{B}\mathtt{B}} y_{\mathtt{B}}
    - \diag(x_\mathtt{B}) D_{\mathtt{B}\mathtt{B}} x_{\mathtt{B}}.
    \label{eq:KKT-bdry}
\end{align}
Consequently, the disk conformal map can be obtained via the following steps.
\begin{itemize}
    \item[1)] Find the minimizer $\theta^*$ of the conformal energy in \eqref{eq:EC_discrete2} by the gradient-type algorithms, such as the gradient descent method and the quasi-Newton method, with the given gradient in \eqref{eq:KKT-bdry}.
    \item[2)] Calculate the boundary Cartesian coordinates $[x_\mathtt{B}, y_\mathtt{B}]$ by \eqref{eq:polar}. Then solve the linear system in \eqref{eq:KKT-int} to obtain the interior vertices $[x_\mathtt{I}, y_\mathtt{I}]$.
\end{itemize}

\section{Average $\mathbf{G}_{\mathbf{0}}(\mathbf{r}, \mathbf{r}^\prime)$ over a triangular element}\label{B:Griri}
\setcounter{equation}{0}  
\renewcommand{\theequation}{B.\arabic{equation}}
Here we give the derivation of the average term $G_{0}(\mathbf{r}, \mathbf{r}^\prime)$ over the each triangular element in the shapes we mentioned before. Considering the negligible size of the triangular elements, $G_{0}$ can be reduced to $1/\left(4\pi\epsilon_0\left|\mathbf{r}-\mathbf{r}^{\prime}\right|\right)$. Thus the average for $4\pi\epsilon_0G_{0}(\mathbf{r}, \mathbf{r}^\prime)$ over a triangle given by
\begin{equation}
    I=\frac{ \int \int d \mathbf{r} d \mathbf{r}^\prime }{\Delta \sigma^2 \left|\mathbf{r}-\mathbf{r}^{\prime}\right|},
\end{equation}
 where $\Delta \sigma$ is the area of the triangle with verticies $\mathbf{r}_{a, b, c}$. In barycentric coordinates the positions are $\mathbf{r}=\mathbf{r}_{a}+t_{1} \mathbf{r}_{b a}+t_{2} \mathbf{r}_{c a}$ where $0<t_i<1, i=1,2, \mathbf{r}_{b a}=\mathbf{r}_{b}-\mathbf{r}_{a}$. We get
\begin{equation}
  I=\frac{J^{2} \int_{0}^{1} \int_{0}^{1} \int_{0}^{1} \int_{0}^{1} d t_{1} d t_{2} d t_{1}^{\prime} d t_{2}^{\prime}}{\Delta \sigma^2 |\left(t_{1}-t_{1}^{\prime}\right) \mathbf{r}_{b a}+\left(t_{2}-t_{2}^{\prime}\right) \mathbf{r}_{c a} |}.
\end{equation}
Here, the Jacobian $J=x_{b a} y_{c a}-x_{c a} y_{b a}$; The area $\Delta \sigma=0.5 |J|$. Now change coordinates to $u_{i}=t_{i}-t_{i}^{\prime}, v_{i}=t_{i}+t_{i}^{\prime}$. The Jacobian for this is equal to $1 / 4$. The denominator of $I$ depends only on $u_{i}$ with a range $-1<u_{i}<1$. For a given $u_{i}$, the integration over $v_{i}$ gives a factor $2-2\left|u_{i}\right|$. We thus get
\begin{equation}
I=\frac{ \int_{-1}^{1} \int_{-1}^{1} d u_{1} d u_{2}\left(1-\left|u_{1}\right|\right)\left(1-\left|u_{2}\right|\right)}{\left|u_{1} \mathbf{r}_{b a}+u_{2} \mathbf{r}_{c a}\right|}.
\end{equation}
Without loss of generality let us take $r_{c a}/r_{b a}=s>1$. We get
\begin{equation}
I=\frac{1}{r_{b a}} \frac{\int_{-1}^{1} \int_{-1}^{1} d u_{1} d u_{2}\left(1-\left|u_{1}\right|\right)\left(1-\left|u_{2}\right|\right)}{\left|u_{1} \mathbf{e}_{ba}+u_{2} s \mathbf{e}_{ca}\right|}
\end{equation}
here the unit vectors $\mathbf{e}_{ba}=\mathbf{r}_{b a}/r_{b a}$ and $\mathbf{e}_{ca}=\mathbf{r}_{c a}/r_{c a}$. The integral 
$I$ is split into four regions:
\begin{equation}
    I = I_{++} + I_{+-} + I_{-+} + I_{--},
\end{equation}
where subscripts $\pm$ denote the sign of $u_1$ and $u_2$ over $[-1,1]$. 

For example, $I_{-+}$ corresponds to $u_1\in[-1,0]$ and $u_2\in[0,1]$ 
\begin{equation}
I_{-+}=\frac{1}{r_{b a}} \frac{\int_{-1}^{0} d u_{1} \int_{0}^{1} d u_{2}\left(1+u_{1}\right)\left(1-u_{2}\right)}{\left|u_{1} \mathbf{e}_{ba}+u_{2} s \mathbf{e}_{ca}\right|}.
\end{equation}
Define $w_{1}=-u_{1}$, we get
\begin{equation}
I_{-+}=\frac{1}{r_{b a}} \frac{\int_{0}^{1} d w_{1} \int_{0}^{1} d u_{2}\left(1-w_{1}\right)\left(1-u_{2}\right)}{\left|-w_{1} \mathbf{e}_{ba}+u_{2} s \mathbf{e}_{ca}\right|}.
\end{equation}
Using the formula $1 /\left|r-r^{\prime}\right|=\sum_{l} r_{<}^{l} / r_{>}^{l+1} P_{l}(\cos \theta)$, we get
\begin{equation}
I_{-+}=\frac{1}{r_{b a}} \sum_{l} P_{l}(\cos \theta) \int_{0}^{1} d w_{1}\left(1-w_{1}\right)\left[\int_{0}^{\frac{w_{1}}{s}} \frac{d u_{2}\left(1-u_{2}\right)\left(u_{2} s\right)^{l}}{w_{1}^{l+1}}\right. \\
\left.+\int_{\frac{w_{1}}{s}}^{1} \frac{d u_{2}\left(1-u_{2}\right) w_{1}^{l}}{\left(u_{2} s\right)^{l+1}}\right],
\end{equation}
here $\cos \theta=\mathbf{e}_{ba}\cdot \mathbf{e}_{ca}$. The term with $l=1$ is
\begin{equation}
I_{-+}=\frac{1}{r_{b a}} P_{1}(\cos \theta) \int_{0}^{1} d w_{1}\left(1-w_{1}\right)\left[\int_{0}^{\frac{w_{1}}{s}} \frac{d u_{2}\left(1-u_{2}\right)\left(u_{2} s\right)}{w_{1}^{2}}+\int_{\frac{w_{1}}{s}}^{1} \frac{d u_{2}\left(1-u_{2}\right) w_{1}}{\left(u_{2} s\right)^{2}}\right].
\end{equation}
The total integral contribution is
\begin{equation}
\int_{0}^{1} d w_{1}\left(1-w_{1}\right)\left[\frac{3}{2s} - \frac{4w_1}{3s^2} + \frac{w_1}{s^2} \ln \frac{w_1}{s}\right].
\end{equation}
This is finite. Performing the integration over $u_{2}$, the other terms are
\begin{align}
I_{-+} = \frac{1}{r_{b a}} \sum_{l} P_{l}(\cos \theta) \int_{0}^{1} d w_{1}\left(1-w_{1}\right)\left[\frac{2l + 1}{s \, l(l+1)} - \frac{w_1 (2l + 1)}{s^2 (l+2)(l-1)} + \frac{w_1^l}{s^{l+1} \, l(l-1)}\right].
\end{align}
The term proportional to $1/(l-1)$ for $l=1$ corresponds to the term discussed above and is finite. Performing the integration over $w_{1}$, we finally obtain
\begin{gather}
I_{-+}=\frac{1}{r_{b a}} \sum_{l} P_{l}(\cos \theta) g_l,
\end{gather}
where 
\begin{gather}
g_l=\frac{2l + 1}{2 s \, l(l + 1)} 
  - \frac{2l + 1}{6 s^2 (l + 2)(l - 1)} 
  + \frac{1}{s^{l + 1} \, l(l - 1)(l + 1)(l + 2)}.
\end{gather}

In the same way, we get
\begin{equation}
I_{+-}=\frac{1}{r_{b a}} \frac{\int_{0}^{1} d u_{1} \int_{0}^{1} d w_{2}\left(1-u_{1}\right)\left(1-w_{2}\right)}{\left|u_{1} \mathbf{e}_{ba}-w_{2} s \mathbf{e}_{ca}\right|}=I_{-+}.
\end{equation}
Similarly, we get
\begin{equation}
I_{++}= \frac{1}{r_{b a}} \frac{\int_{0}^{1} d u_{1} \int_{0}^{1} d u_{2}\left(1-u_{1}\right)\left(1-u_{2}\right)}{\left|u_{1} \mathbf{e}_{ba}+u_{2} s \mathbf{e}_{ca}\right|}= \frac{1}{r_{b a}} \sum_{l} P_{l}(-\cos \theta) g_l,
\end{equation}
\begin{equation}
I_{--}= \frac{1}{r_{b a}} \frac{\int_{0}^{1} d w_{1} \int_{0}^{1} d w_{2}\left(1-w_{1}\right)\left(1-w_{2}\right)}{\left|-w_{1} \mathbf{e}_{ba}-w_{2} s \mathbf{e}_{ca}\right|}=I_{++}.
\end{equation}

Since $P_{l}(-x)=(-1)^l P_{l}(x)$, we get
\begin{equation}
    I=I_{++}+I_{--}+I_{+-}+I_{-+}=\frac{4}{r_{b a}} \sum_{l} P_{2l}(\cos \theta) g_{2l}.
\end{equation}

\acknowledgements{T. Li was supported by the National Natural Science Foundation of China (grant no. 12271377) and the Jiangsu Provincial Scientific REsearch Center of Applied Mathematics  (grant no. BK20233002). This research was partially funded by Shanghai Institute for Mathematics and Interdisciplinary Science (grant no. SIMIS-ID-2024-LG).}

 \section*{Author Contributions}
 S. T. C., T. L., and S. T. Y. conceived the basic idea of the work and supervised its
development.  P. W. and Z. H. T. performed theoretical and numerical calculations. S. T. C. and P. W. analyzed the computation results. P. W. and Z. H. T. wrote the paper. All authors took part in discussions and revisions of the manuscript.

\section*{COMPETING INTERESTS}
 The authors declare no competing interests.

\bibliographystyle{unsrt}
\bibliography{ccmp}

@book{knott2004radar,
  title={Radar cross section},
  author={Knott, Eugene F and Schaeffer, John F and Tulley, Michael T},
  year={2004},
  publisher={SciTech Publishing}
}

@article{gu2008computational,
  title={Computational conformal geometry},
  author={Gu, Xianfeng David and Yau, Shing Tung},
  journal={(No Title)},
  year={2008}
}

@book{chui2012electromagnetic,
  title={Electromagnetic behaviour of metallic wire structures},
  author={Chui, ST and Zhou, Lei},
  year={2012},
  publisher={Springer Science \& Business Media}
}

@article{chui2014resonances,
  title={Resonances and circuit theory for the interaction of metallic disks and annuli with an electromagnetic field},
  author={Chui, ST and Du, JJ and Yau, ST},
  journal={Physical Review E},
  volume={90},
  number={5},
  pages={053202},
  year={2014},
  publisher={APS}
}

@article{cone1,
 title={Scattering of electromagnetic waves from a cone with conformal mapping: Application to scanning near-field optical  microscope},
  author={S. T. Chui and Xinzhong Chen and Mengkun Liu and Zhifang Lin and Jian Zi },
  journal={Physical Review B},
  volume={97},
  pages={081406 (R)},
  year={2018},
  publisher={APS}
}

@article{pendry2000negative,
  title={Negative refraction makes a perfect lens},
  author={Pendry, John Brian},
  journal={Physical review letters},
  volume={85},
  number={18},
  pages={3966},
  year={2000},
  publisher={APS}
}

@article{smith2004metamaterials,
  title={Metamaterials and negative refractive index},
  author={Smith, David R and Pendry, John B and Wiltshire, Mike CK},
  journal={science},
  volume={305},
  number={5685},
  pages={788--792},
  year={2004},
  publisher={American Association for the Advancement of Science}
}

@article{chen2006active,
  title={Active terahertz metamaterial devices},
  author={Chen, Hou-Tong and Padilla, Willie J and Zide, Joshua MO and Gossard, Arthur C and Taylor, Antoinette J and Averitt, Richard D},
  journal={Nature},
  volume={444},
  number={7119},
  pages={597--600},
  year={2006},
  publisher={Nature Publishing Group UK London}
}

@article{zhou2006eigenmodes,
  title={Eigenmodes of metallic ring systems: A rigorous approach},
  author={Zhou, Lei and Chui, ST},
  journal={Physical Review B—Condensed Matter and Materials Physics},
  volume={74},
  number={3},
  pages={035419},
  year={2006},
  publisher={APS}
}

@article{zhan2014t,
  title={t matrix of metallic wire structures},
  author={Zhan, TR and Chui, ST},
  journal={Journal of Applied Physics},
  volume={115},
  number={14},
  year={2014},
  publisher={AIP Publishing}
}

@article{zhan2015multiple,
  title={Multiple scattering of metallic wire structures},
  author={Zhan, TR and Chui, ST and Lin, ZF},
  journal={Journal of Applied Physics},
  volume={118},
  number={16},
  year={2015},
  publisher={AIP Publishing}
}

@article{chui2016scattering,
  title={Scattering of electromagnetic waves from surfaces with conformal mapping: An example of a triangular plate},
  author={Chui, ST and Wang, Shubo and Chan, Che Ting},
  journal={Physical Review E},
  volume={93},
  number={3},
  pages={033302},
  year={2016},
  publisher={APS}
}

@article{yueh2017efficient,
  title={An efficient energy minimization for conformal parameterizations},
  author={Yueh, Mei-Heng and Lin, Wen-Wei and Wu, Chin-Tien and Yau, Shing-Tung},
  journal={Journal of Scientific Computing},
  volume={73},
  pages={203--227},
  year={2017},
  publisher={Springer}
}

@article{chui2017electromagnetic,
  title={Electromagnetic sensors from algebraic corner vortex generation in polygonal plates},
  author={Chui, ST and Wang, Shubo and Chan, Che Ting},
  journal={Applied Physics Letters},
  volume={110},
  number={1},
  year={2017},
  publisher={AIP Publishing}
}

@article{YCWW21,
  author = {Yueh-Cheng Kuo and Wen-Wei Lin and Mei-Heng Yueh and Shing-Tung Yau},
  title = {Convergent Conformal Energy Minimization for the Computation of Disk Parameterizations},
  journal = {SIAM Journal on Imaging Sciences},
  volume = {14},
  number = {4},
  pages = {1790--1815},
  year = {2021},
  doi = {10.1137/21M1415443},
}

@article{Hutc91,
  author    = {John E. Hutchinson},
  title     = {Computing conformal maps and minimal surfaces},
  journal   = {Proceedings of the Centre for Mathematics and its Applications},
  year      = {1991},
  volume    = {26},
  pages     = {140--161},
}

@article{yu2011light,
  title={Light propagation with phase discontinuities: generalized laws of reflection and refraction},
  author={Yu, Nanfang and Genevet, Patrice and Kats, Mikhail A and Aieta, Francesco and Tetienne, Jean-Philippe and Capasso, Federico and Gaburro, Zeno},
  journal={science},
  volume={334},
  number={6054},
  pages={333--337},
  year={2011},
  publisher={American Association for the Advancement of Science}
}

@article{yu2014flat,
  title={Flat optics with designer metasurfaces},
  author={Yu, Nanfang and Capasso, Federico},
  journal={Nature materials},
  volume={13},
  number={2},
  pages={139--150},
  year={2014},
  publisher={Nature Publishing Group UK London}
}

@article{brongersma2025second,
  title={The second optical metasurface revolution: moving from science to technology},
  author={Brongersma, Mark L and Pala, Ragip A and Altug, Hatice and Capasso, Federico and Chen, Wei Ting and Majumdar, Arka and Atwater, Harry A},
  journal={Nature Reviews Electrical Engineering},
  pages={1--19},
  year={2025},
  publisher={Nature Publishing Group}
}

\end{document}